\documentstyle [12pt,cite,subeqn]{article}
\title{Nonrelativistic Quantum Particle in a Curved Space as a Constrained
System\thanks{Supported in part by Conselho Nacional de
Desenvolvimento Cient\'{\i}fico e Tecnol\'ogico (CNPq) and Funda\c{c}\~ao de
Amparo \`a Pesquisa do Estado do Rio Grande do Sul (FAPERGS), Brazil.}}

\author{A. Foerster, H. O. Girotti and P. S. Kuhn \\Instituto de F\'{\i}sica,
Universidade Federal do Rio Grande do Sul \\ Caixa Postal 15051, 91501-970  -
Porto Alegre, RS, Brazil.}

\date{September 1994}

\begin{document}
\maketitle

\begin{abstract}

The operator and the functional formulations of the dynamics of constrained
systems are explored for determining unambiguously the quantum Hamiltonian of a
nonrelativistic particle in a curved space.

\end{abstract}
\end{document}

\documentstyle[preprint,aps]{revtex}
\begin{document}
\setcounter{page}{2}

The canonical quantization of a nonrelativistic particle moving
in a curved space is a long standing problem
\cite{DW1,DW2,Ma}. Indeed, for such system the
classical-quantum correspondence does not define a unique
Hamiltonian operator. The free massive particle already serves to
illustrate this point; the corresponding classical
Hamiltonian is\footnote{The index $i$ runs from 1 to N-1, $p_{i}$
is the momentum canonically conjugate to the particle coordinate
$q^{i}$ and $g^{ij}$ denotes the inverse
of the metric tensor $g_{ij}$. Everywhere in this paper repeated indices
sum over their corresponding ranges.} $h =
\frac{1}{2M}p_{i} g^{ij}(q) p_{j}$, implying that the classical-quantum
transition $\,h  \rightarrow H\,$  is afflicted by ordering ambiguities.
Moreover, not all quantum mechanical counterparts of $\, h \,$
possess a coordinate representation
behaving as scalar under generalized coordinate transformations
\footnote{By assumption, the wave function of a spinless particle
$\phi(q) = <q \mid  \phi >$ behaves as scalar under generalized coordinate
transformations.}\cite{DW1,DW2,Ma}. The canonical quantization
procedure is, then, plagued with ambiguities which are not
harmless, since they affect the energy spectrum of the physical
system\cite{Kl1}.

The outcomes from the path-integral approach can be summarized as
follows\cite{DW2,Ma,Kl1,Kl2,Ch,De}

\begin{equation}
<q \mid H \mid \phi>
\, = \, - \frac {\hbar^ {2}} {2M \sqrt{g}} \partial_{i}
\left[ \sqrt{g} g^{ij} \partial_{j} \phi(q) \right] \,\,
+ \,\, \frac {\kappa \hbar^ {2}}{M} R \phi(q) \,\, ,
\label{1}
\end{equation}

\noindent
where $g = \det g_{ij}$ and $\partial_i \equiv \partial/\partial q^{i}$. As
can be seen, besides the Laplace-Beltrami operator, a term proportional to the
scalar curvature $R$ arises in the right hand side of (1). For the
dimensionless
constant $\, \kappa \,$, the values found by  DeWitt\cite{DW2}, Cheng\cite{Ch}
and Dekker\cite{De} are, respectively, $\kappa=1/12$, $\kappa=1/6$ and
$\kappa=1/8$. For spherical
surfaces (\ref{1}) can be contrasted against the corresponding result
obtained from the quantization via the rotational Lie algebra
\cite{Kl1,Kl2}. For agreement, $\kappa=0$ must be set in (\ref{1}).

Therefore, an unambiguous determination of the Hamiltonian operator describing
the quantum dynamics of a free particle moving in a curved space is
still lacking. In this paper we make a proposal pointing towards such
determination. We shall first work out the problem within the operator
approach and then, separetely, within the functional approach.

Our main idea consists in treating the curved manifold as a hypersurface
($U_{N-1}$) embedded in an Euclidean space. To secure that the motion takes
place on $ U_{N-1}$ we let the Cartesian equation of the hypersurface act as a
constraint. The whole problem of formulating the quantum dynamics of
a free particle in a curved space reduces then, essentially, to solve for the
motion of a constrained quantum system in an Euclidean space. Afterwards, we
undo the embedding in order to recover the original problem.

We start by embedding the (N-1)-dimensional curved space into a
N-dimensional Euclidean space $ ( E^{N} ) $ which is spanned by the
Cartesian coordinates $ x^{a} ,\, a = 1,...,N $. The classical dynamics of a
massive particle (mass M) moving freely in $ E^{N} $ is described by the
Lagrangian

\begin{equation}
L_E  =  \frac {1} {2M}{\dot{x}}_a {\dot{x}}^a,
\label{2}
\end{equation}

\noindent
where the dot denotes differentiation with respect to time. For the motion to
take place on
$U_{N-1}$  we add to $L_E$ the equation of this hypersurface ($f(x) = 0$)
through the
Lagrange multiplier $\lambda$\footnote{The replacement of $f(x) = 0$ by
$\dot{f}(x) = 0 $ has been discussed in the literature\cite{Ta}}. The system
under analysis is now a constrained system\cite{Di,Fr,Su,Gi} whose dynamics
is specified by the canonical Hamiltonian

\begin{equation}
h_E  =  \frac {1} {2M}p_a p^a,
\label{3}
\end{equation}

\noindent
together with the second-class constraints

\begin{mathletters}
\label{4}
\begin{eqnarray}
\phi_1 \, & \equiv & \, p_{\lambda} \, \approx \, 0, \label{mlett:a4} \\
\phi_2 \, & \equiv & \, f(x) \approx 0,\label{mlett:b4} \\
\phi_3 \, & \equiv & \, \frac{1}{M} p^a\partial_a f \approx 0,
\label{mlett:c4} \\
\phi_4 \, & \equiv & \, \frac{1}{M} p^a p^b ( \partial_a \partial_b f)
\, + \, \frac{\lambda}{M}\, (\partial^a f) ( \partial_a f) \approx 0.
\label{mlett:d4}
\end{eqnarray}
\end{mathletters}

\noindent
Here, $ p_{\lambda}$ and $p_{a} $ are, respectively, the momenta canonically
conjugate to $\lambda$ and $ x^{a} $ and $\partial_a \equiv
\partial/\partial x^{a}$.

To quantize the system within the operator approach one first promotes the
phase-space variables to Hermitean operators, i.e., $  x^{a}  \rightarrow
X^{a}$ , \,\,$ p_{a} \rightarrow  P_{a} $,
\,\, a  =  1,...,\,N,  \,\,$ \lambda  \rightarrow \Lambda$, \,\,
$p_{\lambda} \rightarrow P_{\Lambda} $,
and then establishes that the $X^{\prime}s  $,  $ P^{\prime}s $, $ \Lambda $
and $ P_{\Lambda} $ obey a set of commutation rules which are abstracted from
the corresponding Dirac brackets\cite{Di}, the constraints thereby translating
into strong operator relations. This is the Dirac bracket quantization
procedure\cite{Di,Fr,Su,Gi} which presently yields

\begin{mathletters}
\label{5}
\begin{eqnarray}
& & \left[ X^{a} \, , \, X^{b} \right] \, = \, 0 ,\label{mlett:a5} \\
& & \left[ X^{a} \, , \, P_{b} \right] \, = \,  i\hbar
\left (\delta^a_b \, - \, n^a n_b \right)\, ,\label{mlett:b5} \\
& & \left[ P_{a} \, , \, P_{b} \right] \,  =  \,  i\hbar
P^c \cdot \left( n_b \partial_c n_a - n_a \partial_c n_b
\right). \label{mlett:c5}
\end{eqnarray}
\end{mathletters}

\noindent
Here, $n^a \equiv \partial_a f/ \sqrt{M \alpha}$ is the normal to the
hypersurface and $ \alpha \equiv (\partial_a f) (\partial^a f)/M$ .
In (\ref{mlett:c5}) the symmetrization prescription
($A \cdot B \equiv (AB+BA)/2$) was called for to guarantee the self-adjointness
of the $P^{\prime}s$.
However, one is to observe that the right hand side of (\ref{mlett:c5}) depends
linearly on the momenta. Hence, all ordering prescriptions compatible with
$P_a = P_a^{\dagger}$ are equivalent, since one can reduce to each other by
means of the commutation relations\cite{DW1}.
To phrase it differently, the need for an ordering prescription in
(\ref{mlett:c5}) does not imply in lack of uniqueness of the classical-quantum
correspondence.

It follows from the above that the composite quantities $ h_E $,
$\phi_2$ and $\phi_3$ can be {\it unambiguously} promoted to the quantum level,
i.e.,

\begin{mathletters}
\label{6}
\begin{eqnarray}
& & h_E \rightarrow H_E  =  \frac {1}{2M} P_a P^a, \label{mlett:a6} \\
& & \phi_2 \rightarrow \Phi_2 =  f(X) = 0, \label{mlett:b6} \\
& & \phi_3 \rightarrow \Phi_3 = \frac{1}{M} P^a \cdot \partial_a f(X) = 0.
\label{mlett:c6}
\end{eqnarray}
\end{mathletters}

\noindent
As for $ \phi_1$ and $\phi_4$, we remark that they just allow for the
elimination of the variables $\Lambda$ and
$ P_{\Lambda}$. As seen from Eqs.(\ref{5}), these variables decouple completely
from the sector of interest ($X, P$). For this reason, we did not include in
(\ref{5}) the commutation relations involving $\Lambda$ and $P_{\Lambda}$.

We shall next look for the physical phase-space, namely, the space spanned by a
set of independent variables obeying canonical commutation
relations\cite{Fr,Fa}. To construct such space, we start by introducing the
operators $Q^{\mu}$, $\Pi_{\mu}$ and $ \bar{g}_{\mu \nu}(Q)$ through the
quantum point transformation

\begin{mathletters}
\label{7}
\begin{eqnarray}
& & Q^{\mu}\,=\,Q^{\mu}(X), \label{mlett:a7} \\
& & \Pi_{\mu}\,=\,e^{a}_{\mu} \cdot P_{a}, \label{mlett:b7} \\
& & \bar{g}_{\mu \nu}\,=\,\bar{g}_{\nu \mu}\,=
\, e^{a}_{\mu} \delta_{a b} e^{b}_{\nu}, \label{mlett:c7}
\end{eqnarray}
\end{mathletters}

\noindent
where the index $\mu$ runs from $0$ to $N-1$, $ \delta_{a b}$ is the Kronecker
delta and $e^{a}_{\mu}\,\equiv \,\partial X^{a} / \partial Q^{\mu}$.
We also choose

\begin{mathletters}
\label{9}
\begin{eqnarray}
& & Q^{0}\,=\,\frac{1}{\sqrt M} \Phi_{2}\,=\,0, \label{mlett:a9} \\
& & \Pi_{0}\,=\,\frac{\sqrt M}{\alpha} \Phi_3\,=\,0. \label{mlett:b9} \\
& & \bar{g}_{0 i}\,=\,0, \label{mlett:c9}
\end{eqnarray}
\end{mathletters}

\noindent
the index $i$ running over the {\it reduced} range  $1 \leq i \leq N-1$.
The operators $Q$, $\Pi$ and $\bar{g}(Q)$ are, respectively, the quantum
analogs of the classical quantities $q$, $\pi$ and $\bar{g}(q)$ resulting from
$x$ and $p$ after the above transformation. Clearly, $q^0$
denotes the coordinate normal to $ U_{N-1} $ whereas the $ q^{i \prime} $s are
the coordinates spanning the hypersurface. On the other hand, $\pi_0$ is the
component of the momentum along the normal, which of course vanishes. Finally,
$ \bar{g}_{0i} = 0 $ incorporates the information that $ q^{0} $ is orthogonal
to the $ q^{i \prime} $s.

In terms of the operators $Q^{\prime}s$ and $\Pi^{\prime}s$, the commutation
relations (\ref{5}) and the Hamiltonian (\ref{mlett:a6}) can be cast as

\begin{mathletters}
\label{10}
\begin{eqnarray}
& & \left[ Q^{i} \, , \, Q^{j} \right] \, =
\,\left[ \Pi_{i} \, , \, \Pi_{j} \right] \,=\, 0 ,\label{mlett:a10} \\
& & \left[ Q^{i} \, , \, \Pi_{j} \right] \, = \,  i\hbar
{\delta}^{i}_{j},  \label{mlett:b10}
\end{eqnarray}
\end{mathletters}

\begin{equation}
\label{11}
H_E\,=\,\frac{1}{2M} g^{-\frac{1}{4}} \Pi_{i} g^{\frac{1}{2}} g^{ij} \Pi_{j}
g^{-\frac{1}{4}}\,+\,V_{Q}\,,
\end{equation}

\noindent
where

\begin{equation}
\label{111}
V_{Q}\,=\,\frac{{\hbar}^{2}}{8M}
\frac{(\partial_{a}f)(\partial^{a}f)}{M}
{\bar{\Gamma}}^{i}_{0i} {\bar{\Gamma}}^{j}_{0j},
\end{equation}

\noindent
${\bar{\Gamma}}^{\mu}_{\nu \rho}$ is the Christoffel symbol\cite{We}
associated with the metric ${\bar{g}}_{\mu \nu}$, and $g$ designates the
{\it reduced} determinant $g \equiv \det{g}_{ij}$.

The first term in
Eq.(\ref{11}) is the standard Laplace-Beltrami operator. It only contains
quantities intrinsic to the surface. On the other hand, $V_{Q}$ represents the
contribution to the Hamiltonian  arising from the quantum fluctuations of the
normal vector. This degree of freedom is not eliminated by the constraint
$Q^{0}\,=\,\Phi_{2}/ \sqrt M \,=\,0$, because derivatives of $Q^0$ appear in
$H_E$. It is easy to see that $V_{Q}$ behaves as scalar under {\it reduced}
coordinate transformations.

If the result (\ref{11}) is interpreted as final\cite{Ko}, one is forced to
conclude that the effect of quantum fluctuations of the normal vector is
unavoidable within the embedding procedure. However, the result (\ref{11})
can not be considered as final since, for recovering the original
system, one must first remove the spurious degree of freedom. This is what we
meant by ``undoing the embedding'' in the opening paragraphs of this note.

To implement the removal above, we start by recalling that the intrinsic
geometry of a surface is based on the inner product as applied only to its
tangent vectors\footnote{see for instance Ref.\cite{Ne}}. In other words, only
tangent vectors to the surface
belong to the calculus of the surface itself. This is certainly not
the case with normal vectors, which enter in the structure of
$V_{Q}$ as $ (\partial_{a}f) (\partial^{a}f) $.

Hence, by isolating
from $H_E$ the piece intrinsic to the surface one arrives to

\begin{equation}
\label{12}
H\, = \,\frac{1}{2M} g^{-\frac{1}{4}} \Pi_{i} g^{\frac{1}{2}} g^{ij} \Pi_{j}
g^{-\frac{1}{4}}.
\end{equation}

\noindent
Thus no term proportional to the scalar curvature shows up in
$H$. This result for the Hamiltonian operator of a
free particle in a curved space is free of ambiguities and coincides
with that found by Kleinert\cite{Kl1,Kl2} through another method.

We turn now into implementing the embedding within the functional
framework. For a constrained system, involving
only second-class constraints, the Feynman kernel $K$ is given by the
Faddeev-Senjanovic\cite{FS} phase-space path-integral

\begin{eqnarray}
\label{13}
K  & = & {(2 \pi \hbar)}^{-(N-1)(m+1)}\,\int
\left\{ \prod_{j=1}^{m} d \vec{x}(j)\right\}\,
\left\{ \prod_{j=0}^{m} d \vec{p}(j)\right\}\,
\left\{ \prod_{j=0}^{m} d\lambda(j)\right\}\,
\left\{ \prod_{j=0}^{m} {dp}_{\lambda}(j)\right\} \nonumber \\
& & \left\{ \prod_{j=0}^{m} \delta \left( p_{\lambda}(j) \right) \right\}
\left\{ \prod_{j=0}^{m} \delta \left( f(\vec{x}(j)) \right) \right\} \,
\left\{ \prod_{j=0}^{m} \delta \left( \frac{1}{M}p_a(j)\partial^a
f(\vec{x}(j)) \right) \right\} \nonumber \\
& & \left\{ \prod_{j=0}^{m} \delta \left( \lambda(j) \alpha ( \vec{x}(j) ) \,
+ \, \frac{1}{M} p_a(j) p_b(j) \partial^a \partial^b f(\vec{x}(j))
\right) \right\}
\left\{ \prod_{j=0}^{m} {\alpha}^2( \vec{x}(j)) \right\} \nonumber \\
& & \exp { \left\{ \frac{i \epsilon}{\hbar} \sum_{j=0}^m \left[ p_a(j)
\, \frac{x^a(j+1) - x^a(j)}{\epsilon} \, - \, \frac{1}{2M} p_a(j) p^a(j) \,
+ \, \lambda(j) f(\vec{x}(j)) \right] \right\} },
\end{eqnarray}

\noindent
where the time-slicing definition has been adopted\cite{Fu} and

\[
d \vec{x} \equiv \prod_{a=1}^N dx^{a}, \,\,\,\,\,\,
d \vec{p} \equiv \prod_{a=1}^N dp_{a}.
\]

\noindent
Furthermore, we have substituted in (\ref{13}) the value of the determinant of
the second-class constraints (see Eqs.(\ref{4})),

\[
\det [\phi_r\, , \, \phi_s] = \alpha^ 4,\,r,s = 1,...,4.
\]

\noindent
By performing the functional integrations in $ \lambda, \, p_\lambda $ and
$ \vec{p} $ one arrives at

\begin{eqnarray}
\label{14}
K  & = & {\left( \frac{M}{2 \pi \hbar i \epsilon}\right) }
^{ \frac{(N-1)(m+1)}{2} } M^{\frac{(m+1)}{2}} \,
\int \left\{ \prod_{j=1}^{m} d \vec{x}(j)\right\}\,
\left\{ \prod_{j=0}^{m} \delta \left( f(\vec{x}(j)) \right) \right\} \,
\left\{ \prod_{j=0}^{m} {\alpha}^{\frac{1}{2}} ( \vec{x}(j)) \right\}
\nonumber \\
& . &
\exp { \left\{ \frac{i M}{2 \hbar \epsilon} \sum_{j=0}^m \sum_{a,b=1}^N
\left[ x^a(j+1) - x^a(j) \right] \Omega_{ab}(\vec{x}(j))
\left[ x^b(j+1) - x^b(j) \right] \right\} },
\end{eqnarray}

\noindent
where $\Omega_{ab}$ is the projection operator ${\Omega}_{ab} \equiv
\delta_{ab} - n_a n_b$.

The idea here consists in exploring Berezin theorem\cite{Be} to read off
the Hamiltonian operator from the corresponding phase-space path-integral. To
implement such strategy we first identify a set of independent variables by
taking advantage of the point transformation $ x, p \longrightarrow q, \pi $,
already used in connection with the operator approach. This allows to cast
Eq.(\ref{14}) as

\begin{eqnarray}
\label{15}
& & K =  { \left( \frac{M}{2 \pi \hbar i \epsilon} \right)}^{
\frac{(N-1)(m+1)}{2} }
\int \left\{ \prod_{j=1}^{m} \prod_{i=1}^{N-1} dq^{i}(j)\right\}
\left\{ \prod_{j=1}^{m} dq^{0}(j) \right \}
\left\{\prod_{j=0}^m \delta \left(q^0 (j)\right) \right\}
\left\{ \prod_{j=1}^{m} g^{\frac{1}{2}}\left (q(j) \right) \right \}
\nonumber \\
& & {\alpha}^{\frac{1}{2}} (0)
\exp { \left\{ \frac{i M}{2 \hbar \epsilon} \sum_{j=0}^m \sum_{a,b=1}^N
\left[ x^a\left(q(j+1)\right) - x^a\left(q(j)\right) \right]
\Omega_{ab}(q(j))
\left[ x^b\left(q(j+1)\right) - x^b\left(q(j)\right) \right] \right\} }.
\end{eqnarray}

After carrying out the $q^0$ integration, we expand all the functions in the
integrand of Eq.(\ref{15}) around the midpoint

\[
\bar{q}(j) \, = \, \frac{q(j+1) \, + \, q(j)}{2}.
\]

\noindent
Having integrated out all constraints, the calculation follows along lines
quite similar to those in Ref.\cite{Ge}. The midpoint choice enables one to
read off the Weyl ordered Hamiltonian operator directly from the, by now,
unconstrained path-integral\cite{Be}. Thus, one gets

\begin{equation}
\label{16}
H^{\prime}_E \, = \, H \, + \, V^{\prime}_Q,
\end{equation}

\noindent
where

\begin{equation}
\label{17}
V^{\prime}_{Q}\,=\,\frac{{\hbar}^{2}}{8M}
\left\{ R + \frac{(\partial_{a}f)(\partial^{a}f)}{4 M}
\left[ 3 g^{ij}_{\phantom{0} ,0} \, g_{ij,0} -
{ \left( g^{ij} \, g_{ij,0} \right) }^2
\right] \right\}.
\end{equation}

\noindent
Again, we undo the embedding by removing from $H^{\prime}_E$ all the terms
which are not intrinsic to the surface. In this way we obtain

\begin{equation}
\label{18}
H^{\prime} \, = \, H \, + \, \frac{{\hbar}^{2}}{8M} R,
\end{equation}

\noindent
as the Hamiltonian operator of a free particle in a curved space.
This result arised unambiguously within the functional approach and
is seen not to coincide with that obtained in the operatorial scheme (see
Eq.(\ref{12})),
the difference being the term proportional to the scalar curvature.
Discrepancies between the operatorial and functional approaches have
been already antecipated in the literature\cite{Ma}.

To summarize, in this paper we have shown how the operator and functional
formulations of the dynamics of constrained systems can be used for obtaining
the quantum Hamiltonian of an unconstrained nonrelativistic particle in a
curved space.

\end{document}